\documentclass{aa}  

\usepackage{graphicx}
\usepackage{txfonts}
\usepackage[breaklinks,colorlinks,urlcolor=blue,citecolor=blue,linkcolor=blue]{hyperref}
\usepackage{multirow,afterpage,pdflscape,lipsum,capt-of,mathtools}

\usepackage[dvipsnames]{xcolor}
\usepackage{txfonts}
\usepackage{soul}
\usepackage{array}

\newcommand{\n}{$\overline{\mathrm{n}}$}

\newcommand{\dnn}{$\mathrm{\delta}$N/N}
\newcommand{\dnsat}{$\mathrm{\delta}$N/N$_\mathrm{sat}$}
\newcommand{\nup}{$\mathrm{\nu_p}$}
\newcommand{\nbar}{$\bar{\mathrm{n}}$}
\newcommand{\Eta}{$\mathrm{\eta}^*$}
\newcommand{\sigsat}{$\mathrm{\sigma_{sat}}$}
\newcommand{\trise}{$<$t$>$}
\newcommand{\Ds}{D$_\mathrm{s}$}
\newcommand{\K}{<$\kappa$>}
\newcommand{\Rsun}{R$_{\odot}$}
\newcommand{\li}{$\mathrm{l_i}$}
\newcommand{\lo}{$\mathrm{l_o}$}
\begin{document}

   \title{Characterising coronal turbulence using snapshot imaging of radio bursts in 80 -- 200\,MHz}
   \author{A. Mohan}

\institute{Rosseland Centre for Solar Physics, University of Oslo, Postboks 1029 Blindern, N-0315 Oslo, Norway\\
    Institute of Theoretical Astrophysics, University of Oslo, Postboks 1029 Blindern, N-0315 Oslo, Norway\\
              \email{atulm@astro.uio.no}
             }
\authorrunning{Mohan et al.}

   \date{Received September 15, 1996; accepted March 16, 1997}
   \date{Received  - ; accepted  - }

 
  \abstract
   {Metrewave solar type-III radio bursts offer a unique means to study the properties of turbulence across coronal heights. Theoretical models have shown that the apparent intensity and size of the burst sources evolve at sub-second scales under the influence of local turbulence. The properties of the evolution vary with { observation} frequency. However, observational studies remained difficult due to the lack of high fidelity imaging capabilities at these fine temporal scales simultaneously across wide spectral bands.}
   {I present a spectroscopic snapshot imaging (0.5\,s, 160\,kHz resolution) study of a type-III burst event across the 80 -- 200\,MHz band. By modelling the temporal variability of the source sizes and intensity at every observation frequency, the characteristics of coronal turbulence are studied across a heliocentric height range of $\approx$ 1.54 -- 1.75 R$_{\odot}$.}
   {To understand the morphological evolution of the type-III source, a 2D Gaussian fitting procedure is used. The observed trends in the source area and integrated flux density are analysed in the framework of theoretical and data-driven models.}
   {The strength of density fluctuations (\dnn) in the corona is derived as a function of height (R). Combined with the archival low frequency data, \dnn\ values across $\approx$ 1.5 -- 2.2\Rsun\ agree within a few factors. The burst decay time ($\tau_\mathrm{decay}$) and the full width at half maximum of the source showed a power-law dependence with frequency, roughly consistent with the results from data-driven models. However, the values of $\tau_\mathrm{decay}$ across frequencies turned out higher than the expected trend. The intrinsic sizes of the burst source were derived, correcting for scatter broadening. This roughly matched the expected size of flux tubes at the coronal heights explored. I also report the observation of an intrinsic anti-phased pulsation in the area and flux density of the source.}
   {}


   \maketitle
%

\section{Introduction}
\label{sec:intro}
Type-III bursts are triggered by supra-thermal electron beams produced at various particle acceleration sites in the solar corona, especially at active regions \citep{Wild1950}. These particle beams move along nearby open magnetic field structures, tracing a range of coronal heights and generating two-stream instability along the trajectory. Langmuir wave turbulence generated by the instability triggers various wave-wave and wave-particle interactions that lead to coherent plasma radiation from the regions along the beam trajectory at the respective local plasma frequencies (\nup) and their harmonics (2\nup) \citep[e.g.][]{ginzburg1958,Tsytovich69, melrose1972}. However, the observed emission is heavily influenced by the stochastic density fluctuations in the medium, primarily owing to the fact that the frequency of the radiation ($\nu$) is close to the local \nup. 

The plasma frequency is a function of local density ($N$), and the local refractive index ($n$) is a strong function of \nup/$\nu$,
\begin{eqnarray}
    \mathrm{\nu_p} &=& \mathrm{\left[8980\left(\frac{N}{1 cm^{-3}}\right)^{0.5}\right] MHz,}\\
    \label{eqn1:omgp}
    \mathrm{n} &=& \mathrm{\sqrt{1-(\nu_p^2/\nu^2)}}.
    \label{eqn2:ref_index}
\end{eqnarray}
So the turbulent fluctuations in the local density ($N$) cause stochastic fluctuations in the local refractive index, which in turn cause the waves to undergo multiple random scatterings across the medium until they propagate to a height beyond which the mean refractive index, \n, is close to 1. 
Several works have studied the wave propagation effects in the corona using Monte Carlo simulations \citep[e.g.][]{steinberg1971,robinson_scat1983}.
A detailed analytical formalism for the radio wave propagation was developed by \cite{Arzner1999} (AM hereafter). Their work provided a set of analytical equations to model the observed temporal profiles of the burst source properties, such as intensity, size, and mean drift in the image plane.
They predicted that the source sizes and intensity would follow a trend of initial rise leading to a maximum value and a subsequent decay. The analytical expressions for these trends encapsulate the basic features of the local coronal turbulence, such as the density fluctuation index (\dnn) { and} the width of the strong scattering region.
However, to apply this model to data and make reliable estimates of the properties of coronal turbulence, a sub-second-scale imaging capability with sufficient imaging fidelity is required. 
The modern interferometric arrays, including the Murchison Widefield Array \citep[MWA;][]{Tingay2013, Wayth18_MWA_phaseII} and the LOw Frequency ARray \citep[LOFAR;][]{vanHaarlem13}, have facilitated such studies.

Using LOFAR observations of a type-III burst, \cite{kontar2017} demonstrated the trend predicted by AM in the source sizes in tandem with intensity and highlighted the relevance of scattering effects in observed burst sources.
\cite{Atul19_typeIIIQPP_turb} presented the temporal evolution profiles for a type-III source associated with a microflare at 0.5\,s resolution observed at various observation frequencies across a 15\,MHz band centred at 118\,MHz.
The authors modelled the initial rise phase of the source size evolution and derived the \dnn\ within the coronal height range sampled by the 15\,MHz band during the event. 
Meanwhile, an extension of the framework from AM was recently carried out by \cite{Kontar19_Arznercopy} (K19 hereafter), who also compared the theoretical model with archival observations. 
The authors presented a detailed simulation framework to also model the wave propagation through both isotropic and anisotropic scattering media.
{ They also derived} power-law functions for the burst source sizes and intensity decay times as a function of frequency using empirical fits to archival data. 
In a recent work, using simulations based on the K19 framework, \cite{Zhang21_SrcSize_dur_pos_turb} performed parametric simulation studies by varying the level of anisotropy and \dnn\ in the scattering medium to predict the position offset, size, and duration of the radio burst sources seen at 35\,MHz.  

However, there has not been much work that has observed and modelled the co-evolution of source morphology and intensity during bursts at sub-second cadence to derive the characteristic parameters of local coronal turbulence, such as \dnn\ or the size of the scattering screen.
The study by \cite{Atul19_typeIIIQPP_turb} was one such study that explored only a small bandwidth of 15\,MHz, which corresponded to only a very small height range (about ten times smaller than the local density scale height), as mentioned by the authors themselves.
The current work aims to perform a similar, but more detailed, analysis across a much larger height range using data taken across a wide spectral band, from 80 -- 240\,MHz, during another type-III event. Data spanning such a wide spectral sampling will help { to understand} the nature of coronal turbulence and its impacts on radio wave propagation across a broad range of heights in the corona.

Section~\ref{sec:obs} presents the details of the solar event and the observations, followed by an analysis in Sect.~\ref{sec:analysis}. A discussion on the inferences from the analysis is presented in Sect.~\ref{sec:discussion}, followed by conclusions.

\section{Observations}
\label{sec:obs}
Archival MWA Phase-I data are used for this work. A search for type-III events observed by MWA in the `picket fence' observation mode was done. In picket fence mode, MWA observes the Sun simultaneously within 12 spectral bands of 2.56\,MHz in width, which are sparsely distributed across the operational bandwidth of 80 -- 240\,MHz. 
From the type-III observations carried out in picket fence mode, the burst sources that satisfy {the following criteria} were shortlisted: (i) the burst region is close to disk centre, (ii) there is no possible flux contamination from transient or persistent bright nearby sources, and (iii) there are temporally separable burst pulses.

The first criterion lets an isotropic scattering model be safely assumed. This is because the coronal open magnetic field structures, along which the type-III initiating supra-thermal electrons travel, are mostly aligned radially. Choosing a disk-centre source ensures the alignment of the line of sight and the radial vector. 
The characteristics of the turbulence power spectrum along the axes perpendicular to the radial vector can typically be assumed to be similar. This lets me assume isotropic scattering in the plane perpendicular to the radial vector (i.e. the image plane for a disk-centre source). 
However, if I choose sources close to the limb, the radial vector is no longer oriented along the line of sight and the shape of the source becomes asymmetric, subject to strong anisotropic scattering effects in the image plane (see AM and K19) due to the different spatial scales of scattering along and across the magnetic field structures. 
These different scales increase the degree of freedom in the problem and make source evolution modelling harder. 
The analytical equations get simpler for the isotropic case and are readily available in AM (and K19). Modelling is also less ambiguous in the isotropic case since one need not worry about the unknown characteristic length scales of open magnetic field structures and instead can use the typical generic scattering scales in the corona \citep[e.g.][]{coles1989,sasi2017}.
The second criterion ensures that the observed radio source evolution is attributed solely to the propagation of radio emission from the compact type-III burst alone and not contaminated by any sources near the burst location. The cases where a pre-existing persistent source was present near or at the burst location were also excluded from the study.
The third criterion directs the search to observations of bursts with pulse-like temporal profiles. Usually in observations of type-III bursts, multiple strokes of bright burst emission can be seen in the dynamic spectra, caused by pulsed particle acceleration episodes \citep[e.g.][]{ash94_pulsdAccl, 2003SoPh..212..401W, Atul19_typeIIIQPP_turb}. 
It is seen in several cases that the time gaps between these pulsed acceleration episodes are shorter than the timescale of the scatter broadening of pulse profiles, making them merge and hard to separate, especially in low frequency observations. 
In such cases, the scatter-broadened source shapes seen in the image plane and the intensity profiles seen in dynamic spectra from multiple episodes would merge, making the individual study of each burst episode hard.
This would make the study of the effect of scattering and local turbulence on an individual burst episode difficult. 
So, by applying these selection criteria to the MWA Phase-I observational database, I plan to identify the type-III events that involve burst episodes that are well separable in the dynamic spectral plane and produced by disk-centred sources distinguishable in the image plane throughout the event. 
These sources will help an unambiguous study be performed on the response of the local corona to the individual burst pulses at sub-second scales across the coronal heights probed by the wide MWA band. 

A search based on the above criteria found only one dataset, from 11 November 2015, 02:29:00 -- 02:34:00\,UT, that was an appropriate match for this work. The observations had a spectro-temporal resolution of 40\,kHz and 0.5\,s. The few other on-disk type-III events were, when imaged, found to not satisfy the second and third criteria.
The event presented here was among the type-III events studied by \cite{Rahman19_typeIIIStats} using archival MWA Phase-I data; the authors explored the evolution of the degree of polarisation, source motion, and intensity decay time as a function of frequency. Source size evolution, crucial to understanding the propagation of radio waves, was not covered in that work. Also, the authors used the full-Sun dynamic spectrum for their studies. In this study, I use spatially resolved dynamic spectra with detailed source structure modelling across the entire frequency and time range of the observations. 
\section{Analysis}
\label{sec:analysis}
  \begin{figure*}
      \centering
      \includegraphics[width=18cm,height=9.5cm]{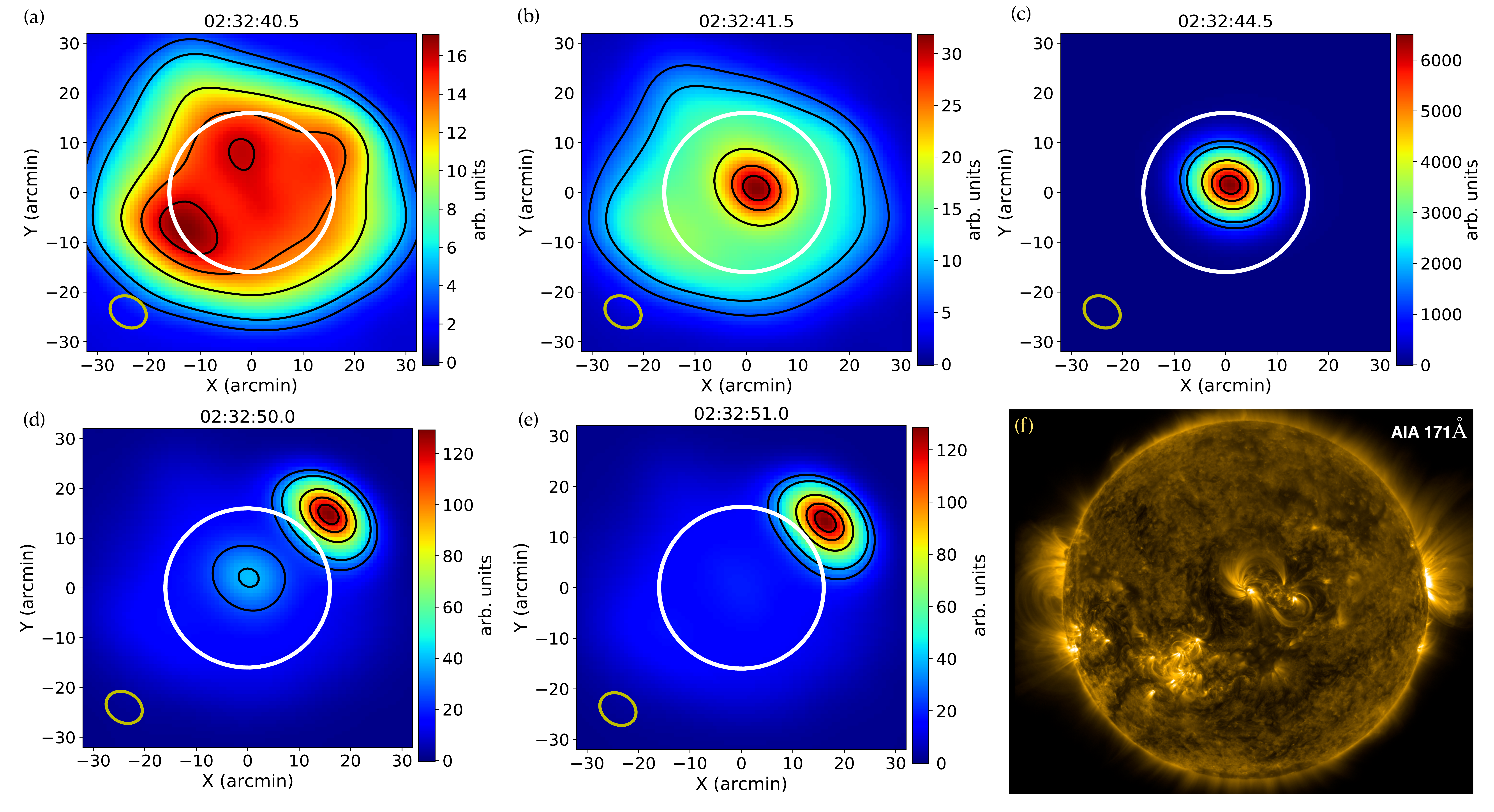}
      \caption{(a--e): 109\,MHz images during different times: (a) just before the start, (b) at the start, (c) at the peak, (d) towards the end, and (e) after the end of the burst. { The images are at the same arbitrary unit scale}. Image contours show 20\%, 30\%, 60\%, 80\%, and 93\% of the peak. The synthesised beam ellipse is shown in the bottom-left corner of the images. The white circle marks the optical disk. (f): AIA 171\AA\ image at 02:32:00\,UT, showing a bright active region at the disk centre during the burst.}
      \label{fig1:burstVsfrq}
  \end{figure*}

Data downloaded from the MWA repository were imaged using the { Automated Imaging Routine
for Compact Arrays for the Radio Sun (AIRCARS)} calibration and imaging pipeline \citep{Mondal_Aircars19}. The absolute calibration of the fluxes were not performed since this is not important for the scientific goals of this study. Thus, the final images are in arbitrary flux units. 
Of the 12 observational bands, the two highest frequency bands had to be discarded owing to bad data quality.
So, the final spectral coverage achieved is 80 -- 200\,MHz.
The burst event started at different times, within 02:32:41.0 -- 02:32:50.5\,UT, in different bands.
Figure~\ref{fig1:burstVsfrq} (a -- e) shows the images of the Sun at the mid-frequency of the operational band during different phases of the burst.  
The burst source that appeared transiently at the disk centre had a 2D Gaussian morphology with no pre-existing sources or nearby bright events before, during, or after the event. The type-III source of interest is possibly related to the active region near the disk centre that is evident in the 171\AA\ image { from the Atmospheric Imaging Assembly (AIA)} shown in Fig.~\ref{fig1:burstVsfrq}(f).
  \begin{figure*}
      \centering
      \includegraphics[width=\textwidth,height=8.5cm]{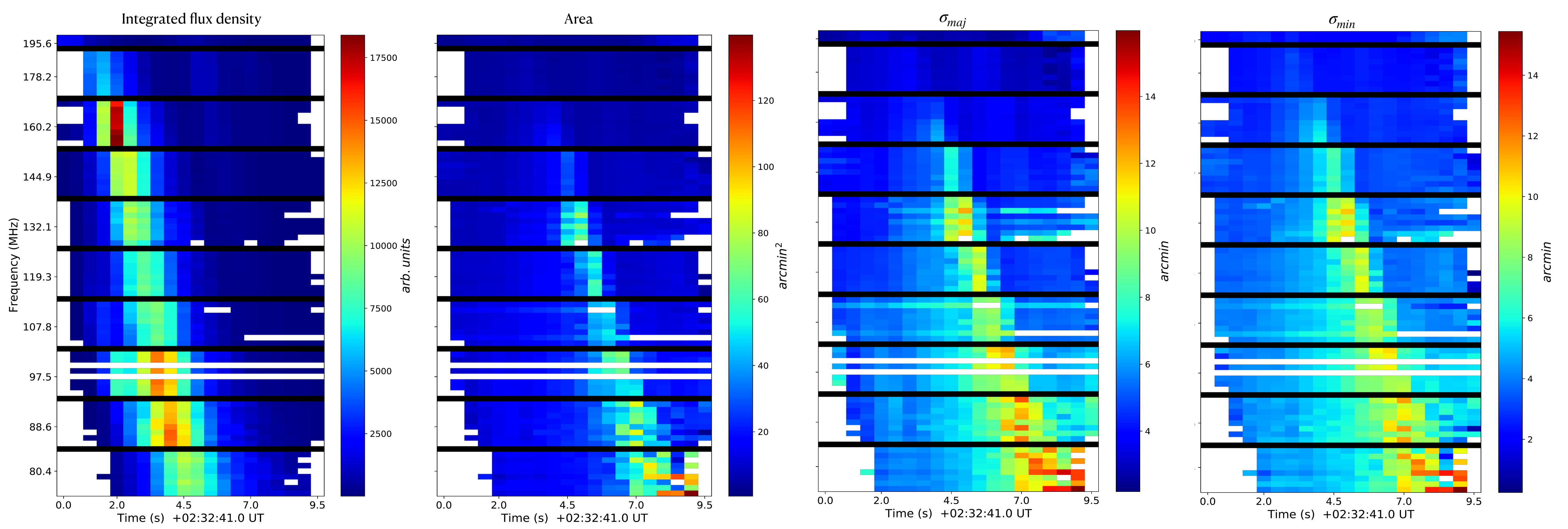}
      \caption{SPREDS for different burst source properties. The black horizontal lines mark the edges of the spectral bands. White regions are masked due to bad data (usually in frequencies above 140\,MHz) or due to the absence of a bright Gaussian burst source.}
      \label{fig2:SPREDS}
  \end{figure*}
\subsection{SPatially REsolved Dynamic Spectrum (SPREDS)}
\label{sec:SPREDS}
The burst source, across all frequencies and times of observation, had a 2D Gaussian morphology. 
To understand the complete spectro-temporal evolution of the burst source, the images across all spectral bands of observation during the entire duration of the event were modelled using the imfit routine in the Common Astronomy Software Applications (CASA; \citealp{casa}) software. 
Each burst image was modelled as a linear combination of a 2D Gaussian function and a constant offset. The offset accounted for the background emission. The parameters of the 2D Gaussian model returned by imfit included the integrated flux density, the peak flux density, the source location, and the full width at half maximum (FWHM) of the major and minor axes of the source after deconvolving the effect of the synthesised beam (FWHM$_\mathrm{maj(min)}$). 
Using FWHM$_\mathrm{maj(min)}$, the standard deviations of the Gaussian source were computed ($\sigma_\mathrm{maj(min)}$), and its area in the image plane was estimated as $\pi\sigma_\mathrm{maj}\sigma_\mathrm{min}$. This definition of area makes it easier to compare the observations with the analytical expressions derived by AM.
The SPatially REsolved Dynamic Spectrum (SPREDS), originally defined for flux density measurements of the source region as a function of time and frequency \citep{Atul17}, has now been extended to other properties of the burst source as well. Figure~\ref{fig2:SPREDS} shows the SPREDS for integrated flux density, area, $\sigma_\mathrm{maj}$, and $\sigma_\mathrm{min}$. The burst happens first in high frequency bands and progresses to lower bands, hinting at a beam of high energy electrons streaming outwards into the corona \citep[see][]{Reid2014}.
Interestingly, the evolutions in $\sigma_\mathrm{maj}$ and $\sigma_\mathrm{min}$ are quite similar, both qualitatively and quantitatively. This supports the practicality of the assumption of isotropic scattering in the image plane. { I note} that the images in the high frequency bands were of poor quality and hence gave no reliable Gaussian fits during the start time of the burst, resulting in the lack of SPREDS data. 

Another interesting feature is the systematic delay of a few seconds in the peaking of the source area with respect to its integrated flux density.
  \begin{figure*}
      \centering
      \includegraphics[width=0.8\textwidth,height=8.5cm]{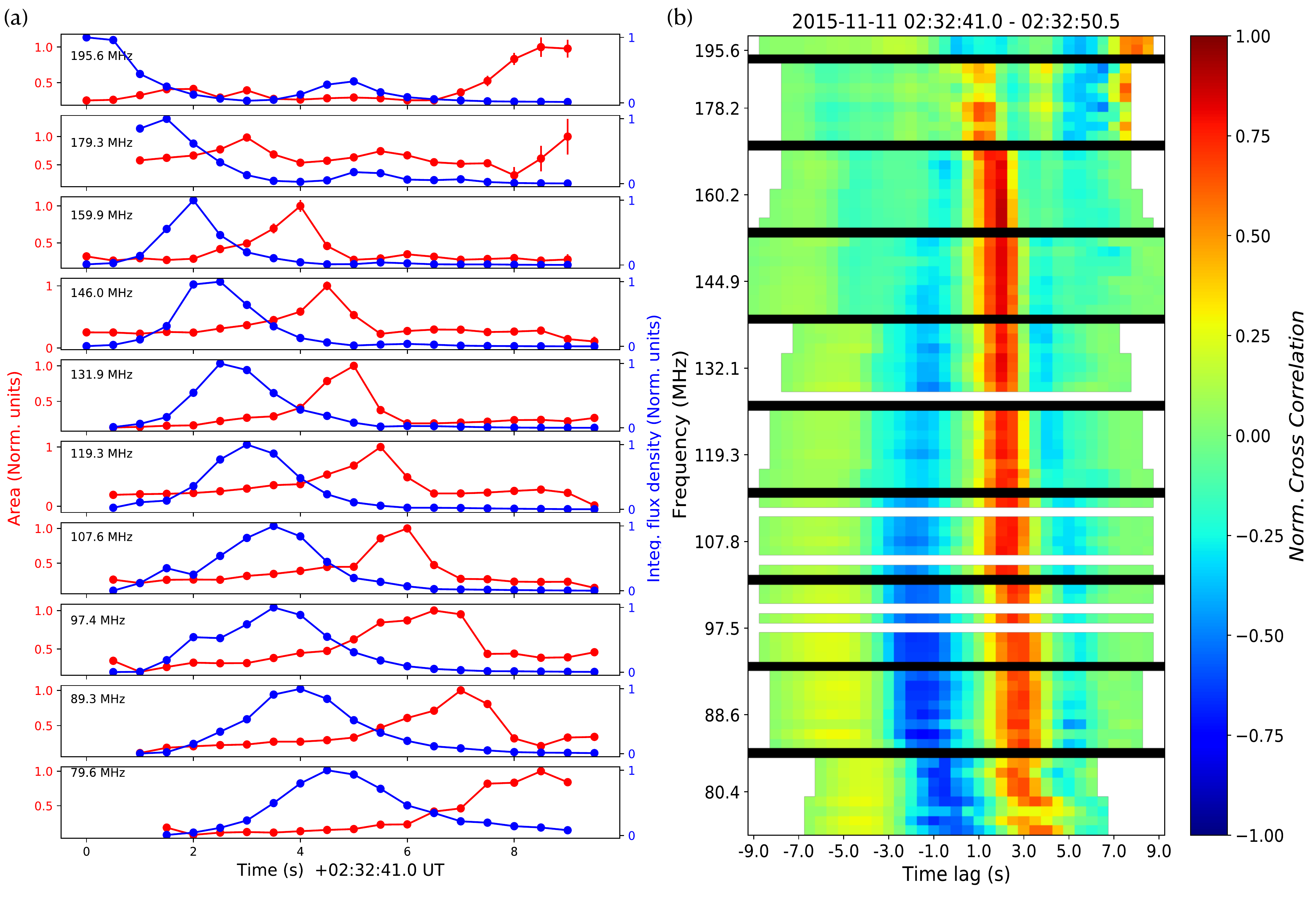}
      \caption{(a): Light curves for area and integrated flux density evolution at the mid-frequencies in each spectral band. Each light curve is normalised by the respective maximum value. (b): NCC analysis for area and integrated flux density.}
      \label{fig3:LC_NCC}
  \end{figure*}
To make this effect clearer, Fig.~\ref{fig3:LC_NCC}(a) shows the mid-band time profiles for area and the integrated flux density for each of the ten picket fence spectral bands. The time profiles are normalised by the respective maximum values to make the presentation clearer without resulting in the distortion or loss of relevant information. The integrated flux density and the area evolution showed pulse profiles in all frequencies except in the two high frequency channels, which suffered from relatively poor imaging quality. The pulse profiles in the two source properties appear shifted in time domain.
Figure~\ref{fig3:LC_NCC}(b) shows the normalised cross-correlation (NCC) analysis done on the two property evolution profiles to robustly estimate this delay in the peaking. I report a delay of $\approx$ 2 -- 3\,s.
Normalised cross-correlation also shows an anti-correlation signature around a time lag of $\approx$ 0 -- 1\,s. However, this, like the correlation around 2\,s, is not a strong signature across all the bands where burst pulses are detected.
A closer look at the type-III events presented by \cite{kontar2017} and \cite{Atul19_typeIIIQPP_turb} also reveals time delays in the peaking of the area and integrated flux density profiles of the same order.
Apart from this, the flux density pulses (hereafter, `burst pulses') tend to broaden as the observation frequency decreases.

\section{Discussion}
\label{sec:discussion}
It is clear from Fig.~\ref{fig2:SPREDS} that the temporal profiles of different source properties are quite similar across the 2.56\,MHz wide spectral bands. The NCC analysis shown in Fig.~\ref{fig3:LC_NCC} also testifies to this. Physically, this is expected since the 2.56\,MHz bandwidth, when converted to the coronal height range, will correspond to less than 5\% of the pressure scale height and density variation scale (N[dN/dR]$^{-1}$) when assuming any of the typical, widely used coronal density models. So the mean properties of the medium explored within this bandwidth are expected to remain similar. Hence, for further analysis, I chose a light curve from every 2.56\,MHz band such that there are fewer masked data points in the time domain and that the observation frequency falls roughly close to the band centre. 
Figure~\ref{fig3:LC_NCC}a shows these chosen light curves in every band with the corresponding observation frequency labelled.
Of these chosen property evolution profiles, I further restricted the analysis to the frequencies below 170\,MHz. This is because the sampling of the burst pulse is poor in the frequencies above 170\,MHz and the objective of the work is to understand the co-evolution of the area and flux density of the burst from the start of the type-III pulse.

To progress further, the observed burst emission was assumed to be produced at the harmonic of the local plasma frequency. This assumption is backed by two facts. Firstly, since the fundamental emission is known to be heavily affected by scattering and other propagation effects \citep[e.g.][]{steinberg1971, robinson_cairns1994, robinson_scat1983}, if only one { stroke} of type-III bursts is observed in the meterwave flux dynamic spectrum, it is most likely to be harmonic emission \citep{Reid2014}. Secondly, the degrees of circular polarisation (p) of the observed bursts were found to be within $\approx$ 10 -- 30\%, with a mean around 20\%, by \cite{Rahman19_typeIIIStats}. This range of values and the mean are below the expected range of p $>30\%$ and mean of p $\approx$35\% for the fundamental emission component \citep{dulk1984}. 
So, using the assumption of harmonic emission, the observation frequencies were converted to densities at the burst site using Eq.~\ref{eqn1:omgp}. Later, using a recent coronal density model by \cite{Alcock18PHD},
\begin{equation}
    \mathrm{N(R)} = \mathrm{4.8\times 10^9\left(\frac{R_\odot}{R}\right)^{14} + 3 \times 10^8\left(\frac{R_\odot}{R}\right)^{6} + 1.4\times10^6\left(\frac{R_\odot}{R}\right)^{2.3}},
    \label{Alcockmodel}
\end{equation}
the heliocentric height of the burst site (R) in the corona was discerned for each observation frequency.
With density and heights estimated, the apparent source properties at each observation frequency could then be modelled to derive the stochastic properties of the corona and the intrinsic burst source dynamics across height. 

\subsection{Effects of scattering in the source evolution}
\label{sec:effects_of_scat}
According to the model for the propagation of radio waves by AM, the wave with frequency $\nu\,<$\,\nup \ undergoes multiple refractions at randomly distributed sites of local over- and under-densities as it propagates outwards towards the observer. This causes the otherwise impulsive narrow time profile of the burst pulse to appear broadened and the apparent source size in the image plane to grow with time. 
The effect of scattering in wave propagation is high when the local refractive index, n, of the wave is considerably less than unity (AM and K19). 
Due to the relatively low n at the generation site, the effect of scattering is more prominent and lasts for a larger extent of the propagation distance for the fundamental emission compared to the harmonic. 
As mentioned in Sect.~\ref{sec:intro}, modern interferometric arrays have facilitated the tracking of the sub-second evolution of the apparent radio burst source structure, subject to radio wave propagation effects across the corona. 
These observations let me then model the propagation effects using an existing theoretical framework and derive the characteristics of coronal turbulence.

\cite{Arzner1999} derived equations relating the scatter-broadened burst intensity and the source area evolution profiles (see Eqs. 62 -- 65). 
In the case of strong scattering, they predicted a linear growth in the observed source area until it reaches a saturation value, which is a function of the strength of density fluctuations, \dnn,\ in the medium.
Meanwhile, the growth rate, \Ds, is a function of the isotropic ray diffusion coefficient, \Eta, which in turn depends on \dnn, the mean inverse spatial ($|\vec{\mathrm{k}}|$) scale (\K), and the spatial profiles of plasma frequency, \nup, and  the refractive index, n, in the medium. The flux density rise time of the burst pulse, \trise, also depends on \Eta\ and \nbar\ during strong scattering.
The authors also defined a `strong turbulence' regime for the scattering medium, during which the \dnn\ in the medium is greater than a theoretical saturation threshold, \dnsat. The \dnsat\ depends only on the basic properties of the scattering medium, namely \nbar\ and its thickness (L). If a scattering medium is in this regime, the saturation value of the apparent source area for an intrinsically point-burst source becomes practically independent of \dnn\ variation in the medium, and it asymptotes to a constant ($\pi$\sigsat$^2$), dependent only on the thickness of the scattering medium. For an intrinsically extended burst source, this saturation area gives the scale of scatter broadening because the effect of the scattering screen on a source is that of a convolution filter.
Equations~\ref{diff} -- \ref{dnsat} present the equations valid for strong scattering derived by AM \citep[also used in][]{Atul19_typeIIIQPP_turb}: 
 \begin{eqnarray}
 \mathrm{D_s} &=& \mathrm{\frac{\pi(c \overline{n})^2}{3\eta^*D^2}}  \label{diff}\\
 \mathrm{<t>} &=& \mathrm{\eta^*\left(\frac{L}{c\overline{n}}\right)^2}   \label{peaktime}\\ 
 \mathrm{\eta^*} &=& \mathrm{\frac{\pi}{8} \frac{c}{\nu^4}<\kappa> 
\overline{\frac{\nu_p^4}{n^3}}\left(\frac{\delta N}{N}\right)^2} \label{eta}\\
\mathrm{\sigma_{sat}} &=& \mathrm{\frac{L}{\sqrt{3}D}}.
    \label{sigma_sat}\\
\mathrm{{\frac{\delta N}{N}}_{sat}} &=& 1.5 \mathrm{\frac{{\overline{n}}^2}{1-{\overline{n}}^2}\left( \frac{l_i}{L} \right)^2}  \label{dnsat}
.\end{eqnarray}
In the above equations, c is the speed of light, D is the Earth-Sun distance, and \li\ is the inner scale length of turbulence. The $\overline{\mathrm{\nu_p^4/n^3}}$ represents the mean of $\mathrm{\nu_p^4/n^3}$ across the thickness of the scattering screen.
The mean density is expected to vary radially outwards across the scattering medium. So, \nup, and thereby n, also varies across the scattering medium as the radio waves propagate outwards. 
It should be noted that since n$\approx$0.86 at the site of generation of the harmonic emission, even a 30\% drop in local density will decrease the effect of scattering significantly as n becomes greater than 0.92. Therefore, the size of the effective scattering screen, L, for the harmonic emission tends to be smaller than the typical local pressure scale height.  
This was demonstrated in a type-III dataset around 111\,MHz by \citet[][their Sect.~\ref{sec:est_scales}]{Atul19_typeIIIQPP_turb}.
Hence, the characteristic properties of the medium, namely \dnn, \K, and \li, are assumed to be { constant within the scattering screen}. 

A closer inspection of the area evolution during the period spanning the rise to decay up to half the maximum value of the burst pulse profile revealed that the area evolution is well defined by a linear growth model (Fig.~\ref{fig3:LC_NCC}(a); see Sect.~\ref{sec:est_scales} for details). 
Such a correlated growth in burst sizes and intensity is expected to result from strong radio wave scattering effects in the corona.
This behaviour can be noted in the data presented by \cite{kontar2017} and \cite{Atul19_typeIIIQPP_turb} as well.
Meanwhile, in the declining phase of the flux density, a more rapid rise in area followed by a dramatic decline, resembling a pulse profile, is seen. 
This resulted in the strong correlation signal seen in the NCC analysis (Fig.~\ref{fig3:LC_NCC}(b)) at a lag of $\approx$ 2 -- 3\,s across all spectral channels with a good sampling of the burst pulse. 
Though an anti-correlation at zero time lag is also found in the NCC plots in some bands, it is noteworthy that this feature is not present in all picket fence bands. 
This is because the area evolution shows a linear growth during the rise of the flux density pulse until its initial declining phase. At higher spectral bands, where the scatter broadening effect is smaller, it becomes clearer that the area pulse and the area rise are distinct events.
A post-burst pulsed evolution in area is not expected from the radio wave propagation models by AM or K19, where the effect of scatter broadening in both area and flux density should happen simultaneously.
So, this pulsation in area could be due to a true variation in the cross-section of the accelerated electron beams that caused the type-III bursts. 
Such time-lagged pulsations in area and integrated flux density at similar delay timescales were reported in the metric type-III bursts by \cite{Atul19_typeIIIQPP_turb}.
This is believed to be caused by the coupling of sausage-like magnetohydrodynamic  modes to the sites of generation of accelerated particle beams that trigger the type-III bursts, as proposed by certain earlier works \citep[e.g.][]{rosenberg1970, Ash2004_SausMode_NthHarmonic_QPO} based on dynamic spectral studies.
Recently, this phenomenon was reported in solar type-I noise storms as well \citep{Atul21_TSmodes, Mondal21_typeI_STmode_survey}.

Since the observed area evolution trends during the type-III burst pulse phase follows a linear trend, characteristic of strong scattering, Eqs. \ref{diff} -- \ref{dnsat} are applicable.
From the linear area growth rate (\Ds) during the burst pulse and the flux density rise time (\trise), the width of the effective scattering layer (L) across the heights probed by various observation frequencies can be found. Assuming the density profile, N(R), the mean refractive index, \nbar, and $\mathrm{\nu_p^4/n^3}$ across L can be computed. 
The estimation of \dnn, however, requires the knowledge of the mean inverse scale length of { turbulence} in the medium, \K.
Assuming Kolmogorov turbulence, the expression for \K\ takes the following form (AM \& K19),
\begin{equation}
    \mathrm{<\kappa>}=\mathrm{2 \frac{{\kappa_o}^{1/3} - {\kappa_i}^{1/3}}{{\kappa_o}^{-2/3} - {\kappa_i}^{-2/3}}},
    \label{eqn9:kappa}
\end{equation}
where $\mathrm{\kappa_i}$ and $\mathrm{\kappa_o}$ are the inner and outer inverse spatial (`k') scales of turbulence in the scattering medium. These k scales are linked to spatial scales as
\begin{eqnarray}
    \mathrm{\kappa_i}&=&\mathrm{2\pi/l_o} \label{eqn10:kappa_i} \\
    \mathrm{\kappa_o}&=&\mathrm{2\pi/l_i} \label{eqn11:kappa_o}
,\end{eqnarray}
where \li\ and \lo\ are the inner and outer spatial scales of turbulence. The value of \li\ is assumed to be three times the local ion inertial scale, based on previous observational results \citep{coles1989, sasi2017}. To get an estimate for \lo, the prescription by \cite{Wohlmuth01_solarTurbscale_lo} was used:
\begin{eqnarray}
    \mathrm{l_i}&=&\mathrm{\left[684\left(\frac{N}{1 cm^{-3}}\right)^{-0.5}\right] km}, \label{eqn12:li}\\
    \mathrm{l_o}&=& \mathrm{\left[0.25 (R/R_\odot)^{0.82}\right] R_\odot}, \label{eqn13:lo}
\end{eqnarray}
where R is the heliocentric height of the radio burst source and N is the local density. 
For a case where $\mathrm{L << l_o}$, $\mathrm{\kappa_i}$ can be practically assumed to be 0. This simplifies Eq.~\ref{eqn9:kappa}, which when recast using Eq.~\ref{eqn11:kappa_o} takes the form
\begin{equation}
    \mathrm{<\kappa>} = \mathrm{4\pi/l_i}
    \label{eqn14:kappa_mean}
.\end{equation}
The values of the different spatial scales (L, \lo, and \li) can thus significantly modify the expression of the mean k scale (\K). This in turn alters the final expression for \Eta, which links the observables to \dnn. So it is important to get reliable estimates for the various spatial scales.

\subsubsection{Estimation of the spatial scales}
\label{sec:est_scales}
The inner (\li) and outer (\lo) scales of turbulence were computed for the observation frequencies by substituting the corresponding N and R estimates in Eqs.~\ref{eqn12:li} and \ref{eqn13:lo}.
The L was estimated as a function of R, using the values for \Ds\ and \trise\ for the respective observation frequencies, as mentioned in the previous section. 
  \begin{figure}
      \centering
      \includegraphics[width=0.45\textwidth,height=6cm]{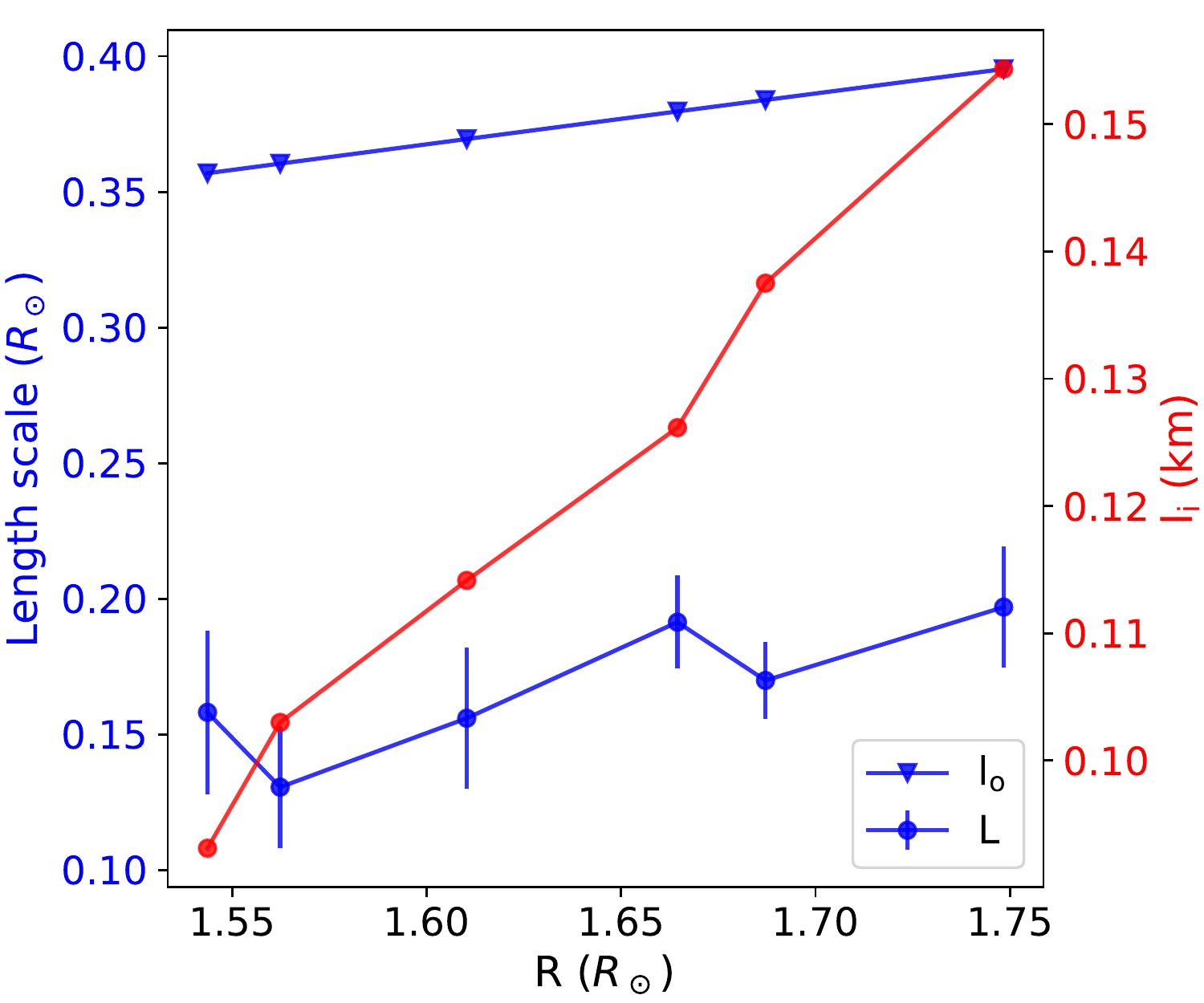}
      \caption{Different scale lengths, L, \li, and \lo , versus R.}
      \label{fig4:L_scales}
  \end{figure}
The \Ds\ was found by fitting a linear function to the area evolution during the flux density pulse period for observation frequencies below 179\,MHz, where the pulse profile is evident in the flux density light curve. 
The $\chi^2$-fitting procedure to deduce \Ds succeeded with errors of less than $\approx$5\% for the area evolution curves at all chosen frequencies except at 159.9 and 146\,MHz. At these two highest of the chosen frequencies, it was found that the area was practically constant during the burst. This could be because the area growth phase was too rapid to be sampled with the time resolution of 0.5\,s. This effect is expected { at} high frequencies. This is because the higher frequencies probe denser and deeper layers of the corona, where the density gradients are higher. Since the effective scattering screen can be envisaged as the region within which the refractive index is considerably less than 1, a steeper local gradient in n leads to a smaller scattering screen width, L. Thinner screens lead to shorter radio wave propagation timescales across them, resulting in faster area growth rates. 
Since the system is under a strong scattering regime, the observed mean area at these frequencies could be the saturation area, \sigsat, predicted by AM.
A corollary to this effect is that the observed burst flux and area pulse profile in frequencies above 146\,MHz should be reflecting a true intrinsic variation since the effect of scatter broadening is already saturated. 
This supports the suggested scenario of sausage-mode-like intrinsic dynamics at the particle { acceleration} site resulting in the time-delayed correlated area-intensity pulsation.
At lower frequency bands, this intrinsic effect is modified by scattering to a larger extent, as is evidenced by the clear initial linear growth of area. 

Figure~\ref{fig4:L_scales} shows the different length scales as a function of heliocentric source heights as derived from the area evolution profiles at the chosen frequencies below 146\,MHz. { The values of \li\ are plotted in km}, while \lo\ and L are in R$_\odot$.
The L values are much less than \lo, which makes Eq.~\ref{eqn14:kappa_mean} a good approximation. Also, L decreases with decreasing R or increasing observation frequencies, as expected.

\subsubsection{Estimating \dnn}
  \begin{figure}
      \centering
      \includegraphics[width=0.4\textwidth,height=6cm]{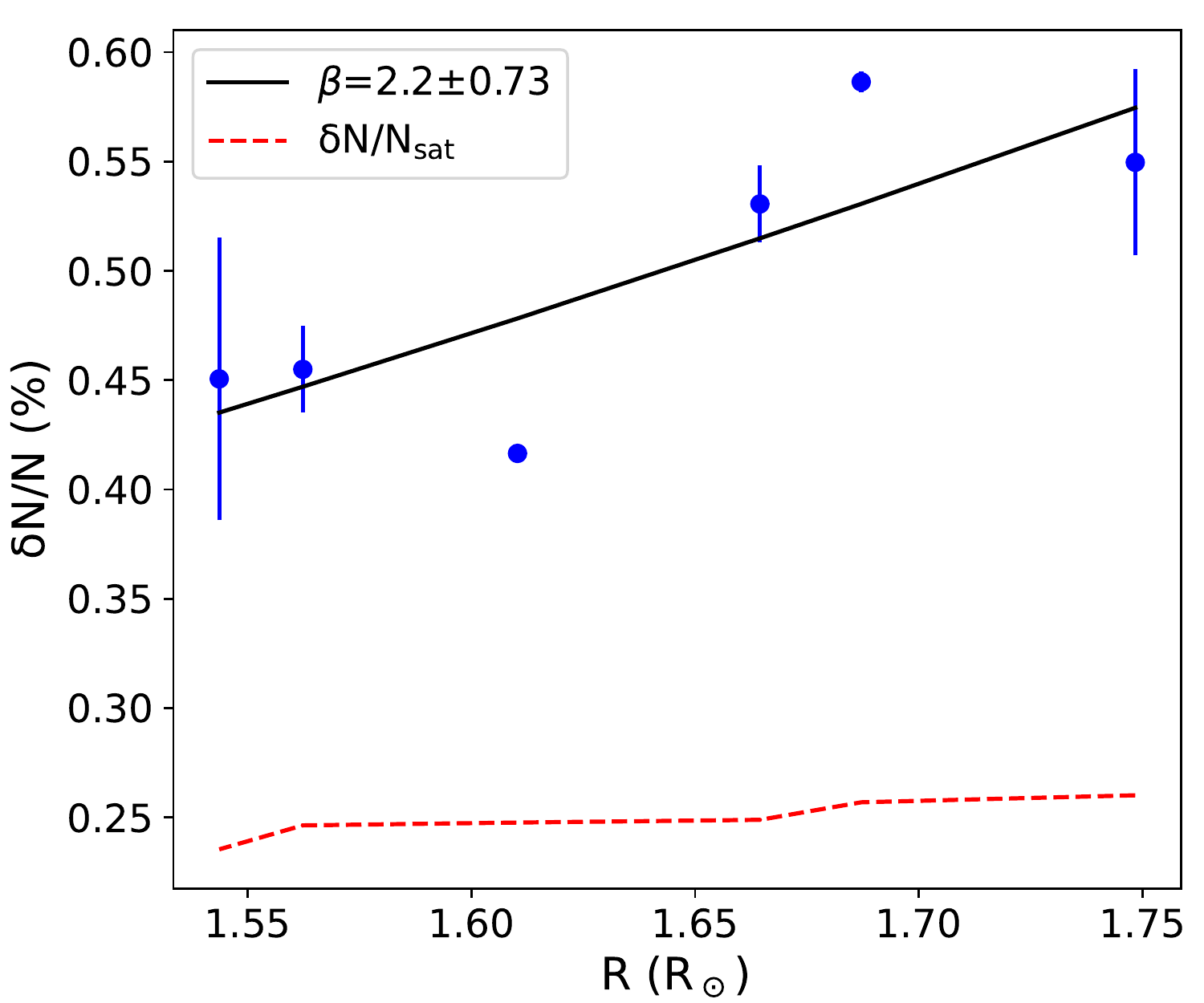}
      \caption{\dnn\ and \dnsat\ as a function of R. A power-law fit to \dnn\ is shown as a solid line.}
      \label{fig5:dnn_vs_ht}
  \end{figure}
With the estimates for L and N across R, \nbar\ and $\overline{\mathrm{\nu_p^4/n^3}}$ were computed. The \Eta(R) was found using Eq.~\ref{diff}. With \K(R) known, and by rearranging Eq.~\ref{eta}, \dnn(R) was obtained. The \dnsat\ was also estimated using Eq.~\ref{dnsat} to know the regime of turbulence -- strong or weak -- as mentioned in Sect.~\ref{sec:effects_of_scat}.
Figure~\ref{fig5:dnn_vs_ht} shows the \dnn(R) and \dnsat(R). The 
\dnn\ values range within 0.42 -- 0.6 \%, with a hint of a rising strength of density fluctuations with height.
A power-law fit assuming \dnn\,$\propto \mathrm{r^\beta}$ gave an index of $\mathrm{\beta}$ = 2.2$\pm$0.73. The
\dnsat \ being greater than \dnn, the system under study is in a strong turbulence regime. { The scatter broadening effect is thus expected to be saturated at a value of \sigsat}. This lets the true intrinsic size of type-III sources be estimated by deconvolving the scatter broadening area scales (\sigsat) from the observed areas across R.

\subsubsection{{Goodness of the assumptions on the mean stochastic properties}}
The framework employed here assumed that the mean properties of the turbulence in the medium, namely \dnn, \li, and \lo,  are fairly uniform across the scattering screen width, L. 
It should be noted that the strength of $\eta*$ can vary within the scattering screen and that it is the variable that reduces the effect of scattering on radio waves as they propagate outwards across L.

Figure~\ref{fig4:L_scales} shows that L $\approx$ 0.15\,\Rsun within $\approx$ 1.52 -- 1.67\,\Rsun. It slowly rises to $\approx$ 0.2  beyond 1.7\,\Rsun. Within the height scale of 0.15 -- 0.2, the \li\ values change only by $<$10\% and \lo\ $\approx$ 5\%. Similarly, \dnn\ is constant within the error bars in the 1.52 -- 1.67\,\Rsun\ and 1.67 -- 1.75\,\Rsun\ ranges. 
Using the derived best-fit model of \dnn, the variational scale of \dnn\ (i.e. [\dnn]/[d\dnn/dR]) $\approx$ R/2.2. This scale is $\approx$\,0.7\Rsun\ at 1.575 and $\approx$\,0.8\Rsun\ at 1.7, which are much higher than L. 
These analyses show that the assumption of uniform stochastic properties within the scattering screen is fairly good and that the derived \dnn\ estimates are reliable.

Another assumption inherent in this model is that the effect of propagation is primarily restricted to within the scattering screen. This is a fair assumption since the scattering is primarily due to the refractive index fluctuations ($\delta n$) induced by the density inhomogeneities in the medium and since the ratio of $\delta n$ to mean local refractive index, $\bar{n}(R)$, is very low outside the scattering screen of width L. The emission being at the harmonic of the local \nup, $\bar{n}$ is already $\sim$ 0.86 at the generation site. By the time it crosses the scattering screen of length L, $\bar{n}\sim$ 0.99. 
From Eq.~\ref{eqn2:ref_index},
\begin{equation}
{\frac{\delta n}{\bar{n}}}= \frac{1}{2\bar{n}^2}\left(\frac{\nu_p}{\nu}\right)^2\frac{\delta N}{N}.    
\end{equation}
At R$\sim$1.65\,\Rsun, assuming a typical L$\sim$0.175\,\Rsun\ and \dnn$\sim$0.5\%, the $\delta n/\bar{n}$ for a harmonic emission originating locally would be $\sim 0.1$\%. But the radiation would experience a $\delta n/\bar{n}$ of only $\sim 0.01$\% at the edge of the scattering screen, L distance away. 
As the wave propagates farther out, the density falls steadily; this makes the $(\nu_p/\nu)^2$ term much smaller and $\bar{n}$ far closer to unity, rendering scattering effects negligible. 
Also, it should be noted that L is a derived quantity based on observed propagation effects, and I find this to be quite small compared to many physical property variation scales. The $\delta n/\bar{n}$ variation and the values discussed justify the smallness of L. 
{ Overall, the model framework is self-consistent.}

\subsubsection{Possible non-scattering effects}
The AM and K19 framework works primarily due to the fact that the observed emission is at the second harmonic, which has a n$\approx$0.86 at the generation site itself, leading to a low value of L.
This high refractive index also ensures that the effects of refraction and reflection on wave propagation will be negligible compared to scattering effects, in contrast to the case for fundamental plasma emission generated close to the n$\approx$0 surface \citep[e.g.][]{Kuznetsov20_radioEcho_driftpair}. However, ensuring the absence of non-scattering effects in the SPREDS data will boost the confidence in the modelling framework used.

The effects of refraction and reflection are usually manifested in the dynamic spectrum as drift-pair (DP) bursts \citep[see ][for an overview]{Melnik05_driftpaibursts}. Drift-pair features appear as short-lived forward or reverse drifting parallel emission streaks imprinted on the type-III emission feature in the dynamic spectrum, with clearly distinguishable drift rates and brightnesses. 
With typical drift rates of $\approx$ 1 -- 2\,MHz/s, they last for $\approx$ 1 -- 2\,s.
Though 0.5\,s resolution can only barely separate any DPs in the time domain, the drift rates of $\sim$\,2\,MHz/s should make the pulse profiles close to the edges of the 2\,MHz band { appear} different. However, the observed pulse profiles are similar across every 2\,MHz band, as seen in the SPREDS (see Fig.~\ref{fig2:SPREDS}).
Additionally, DPs have so far only been reported below 70\,MHz. A long-term survey of these bursts by \cite{delaNoe71_driftpair_survey} showed that 90\% of the observed DPs fell into the 20 -- 45\,MHz range. These events are found mostly associated with type-III storm events, whereas the event in this work is a single isolated burst pulse. 

\subsection{Quantitative comparison with reported scattering effects}
\subsubsection{\dnn}
The \dnn\ has in the past been estimated at different heights in the corona using methods based on, for example, remote sensing \citep{coles1989}, the spectral fine structures in type-III bursts, or the structural deformation of distant radio sources \citep{anantharamaiah1994_VLA_angBroad}. Using the inter-planetary scintillation (IPS) technique \citep[][etc.]{Hewish1955,manoharan1990}, various groups had also estimated the strength of turbulence in the solar wind, which is beyond the scope of this comparative study.

{Meanwhile, by modelling type-IIIb emission features, \citet{sharykin2018_LOFAR_dnn_withtypIIIb} and \citet{mugundhan2017} reported \dnn\ values around 0.1\% and 0.6\% in the 30 -- 80\,MHz band, which explores a height range of $\approx$ 1.7 -- 2.2 R$_\odot$. 
\cite{mugundhan2017} also computed a power-law index, $\beta \approx$\,0.31 $\pm$ 0.1, for \dnn(R), which is quite different from my estimate in the 80 -- 140\,MHz range.
Recently, \cite{Chen20_snapshottypeIIIEvol_turb} computed a mean stochastic fluctuation scale, $\overline{\kappa{(\delta N/N)^2}}$, by applying the K19 model to type-III burst observations at 32\,MHz. Using the reported value and assuming Eq.~\ref{eqn14:kappa_mean}, \dnn\ is estimated as $\approx$0.6\%.   
A caveat in the comparison of these results with each other and with that in the 80 -- 200\,MHz band is the difference in the adopted models for coronal density and characteristic spatial scales, such as \lo\ and \li. 
However, despite these differences in the models and the techniques used, \dnn\ estimates over the wide range from 30 to 140\,MHz (R $\approx$ 1.5 -- 2.2\Rsun) agree within a few factors.}

\subsubsection{Scattered source size}
\begin{figure}
      \centering
      \includegraphics[width=0.35\textwidth,height=9cm]{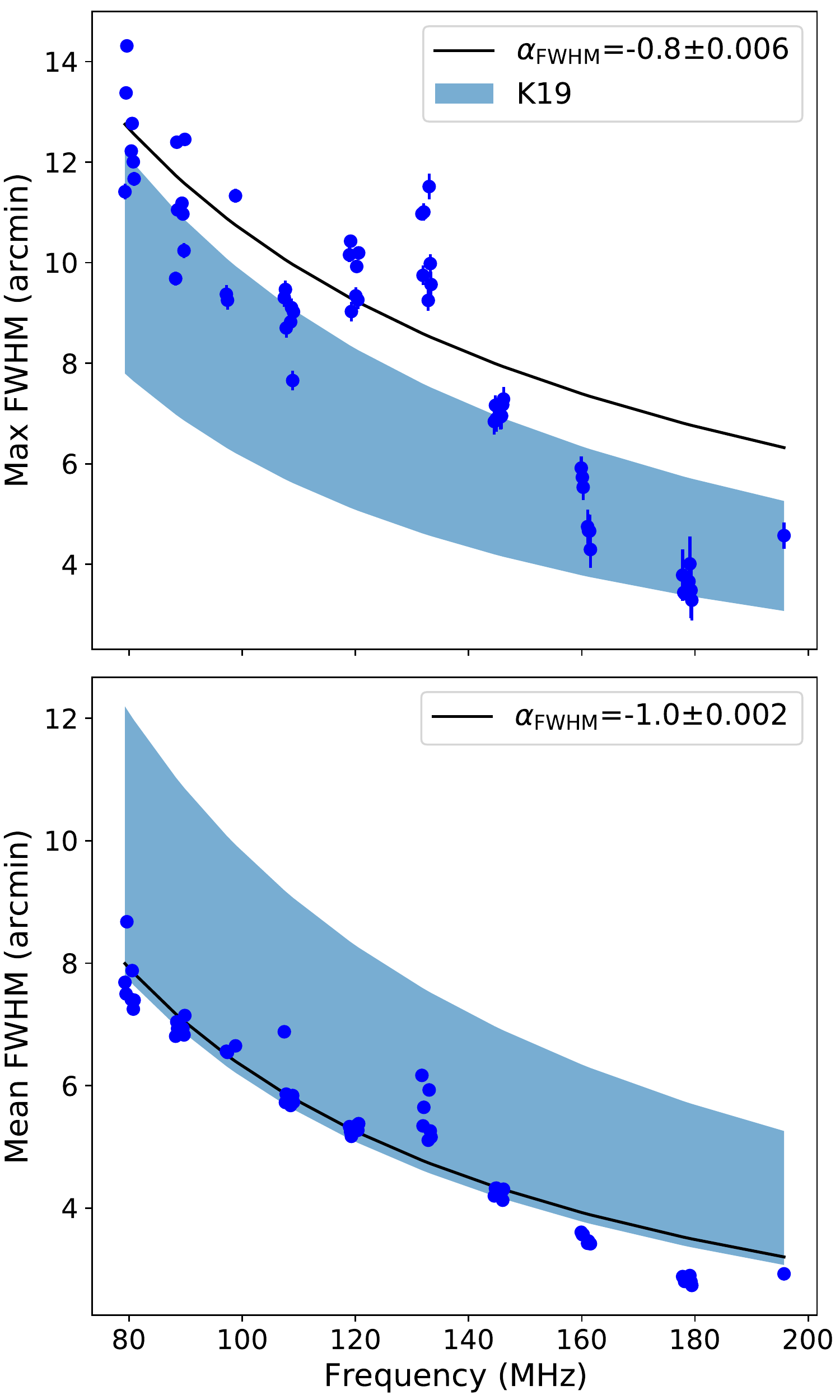}
      \caption{Maximum and mean FWHM observed for the burst source across time at each observation frequency. Solid lines show the power-law fits to the data, and the shaded region shows the expected range of source sizes based on the K19 results.}
      \label{fig6:FWHM_Vs_nu}
  \end{figure}

Owing to the sub-second time resolution of the data, an evolving source size is observed in this work during the type-III event. So, to compare the observed source sizes across frequency with the literature values, which mostly lack the time evolution information, the maximum and the mean FWHM of the source observed over time are considered at each observation frequency. The FWHM is computed as $\sqrt{\mathrm{A_{obs}/\pi}}$, where $\mathrm{A_{obs}}$ is the observed source area. 
A power-law fit was performed to both estimates of the FWHM as a function of frequency, and spectral indices ($\alpha_\mathrm{FWHM}$) were obtained. 
Figure~\ref{fig6:FWHM_Vs_nu} shows the FWHM variation along with the spectral index, $\alpha_\mathrm{FWHM}$, for both mean (-1.0$\pm$0.002) and maximum (-0.8$\pm$0.006) values. The solid curve shows the best-fit function obtained for the data.
The shaded band labelled K19 shows the expected size range from the best-fit model derived by K19 using a compilation of archival estimates of source sizes. It can be seen that the observed mean source size falls roughly within the expected range, but the maximum values tend to fall outside of it, especially at higher frequencies. 
However, if all area estimates across time are over-plotted for every frequency, they would populate the K19 predicted band with some values below and above the predicted range. The $\alpha_\mathrm{FWHM}$ estimates for the mean and maximum area data flank the K19 value of 0.98$\pm$0.05.
  \begin{figure}
      \centering
      \includegraphics[width=0.4\textwidth,height=6cm]{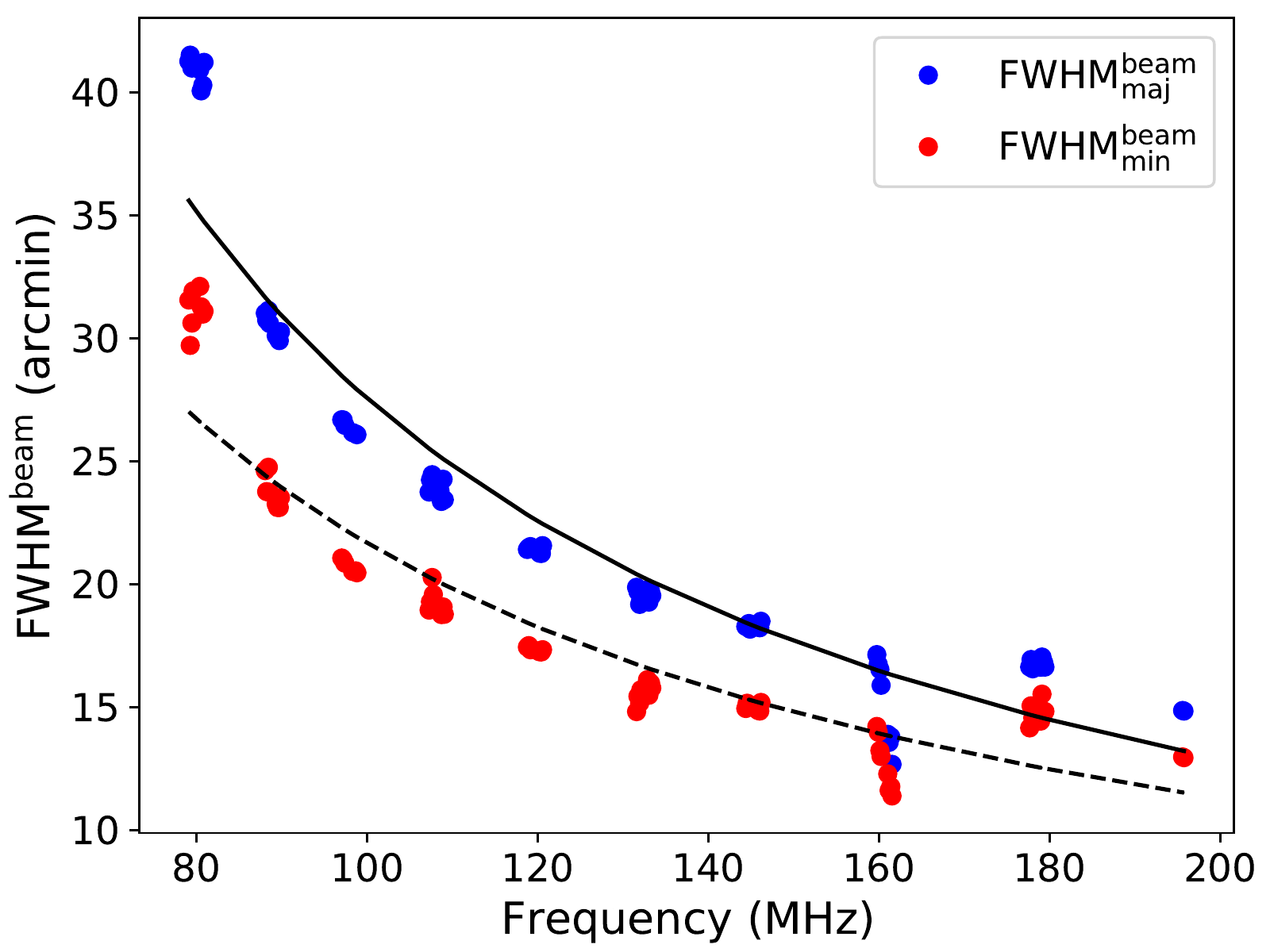}
      \caption{Temporal mean FWHM of the synthesised beam along its major and minor axes. The solid and dashed lines show the best-fit power-law functions for FWHM$^\mathrm{beam}_\mathrm{maj}$ and FWHM$^\mathrm{beam}_\mathrm{min}$, respectively.}
      \label{fig7:FWHM_beamVs_nu}
  \end{figure}

Meanwhile, a significant spike in the observed area is noted at 130\,MHz in both mean and maximum values. 
This is not an anomaly due to variations in the synthesised beam size, since the analysis uses beam-deconvolved source sizes. 
However, to discard the chances of any possible imaging-related systematic effects, the variation in the FWHM of the beam across its major and minor axes (FWHM$^\mathrm{beam}_\mathrm{maj(min)}$) was studied across time and frequency. It was found that the FWHM$^\mathrm{beam}_\mathrm{maj(min)}$ remained constant across time for any particular frequency. The temporal mean of the FWHM$^\mathrm{beam}_\mathrm{maj(min)}$ was estimated as a function of frequency. Figure~\ref{fig7:FWHM_beamVs_nu} shows the spectral variation in this mean FWHM$^\mathrm{beam}_\mathrm{maj(min)}$ along with the best-fit power-law models for the same (solid line: FWHM$^\mathrm{beam}_\mathrm{maj}$; dashed line: FWHM$^\mathrm{beam}_\mathrm{min}$). The spectral indices for FWHM$^\mathrm{beam}_\mathrm{maj}$ and FWHM$^\mathrm{beam}_\mathrm{min}$ were found to be -1 and -0.95, respectively, consistent with the expected trend in angular resolution with frequency for a fixed array configuration. 
Also, no anomalous behaviour is seen in FWHM$^\mathrm{beam}_{maj,min}$ near 130\,MHz. 
So the observed spike in the deconvolved source sizes around 130\,MHz is real.
This frequency corresponds to the height around 1.65 -- 1.7\,R$_\odot$. The \dnn(R), which depends on the growth rate rather than on the absolute value of source area, also showed a local spike in this height range. 
Both these observations suggest an enhanced turbulence in this narrow region.

\subsubsection{Burst pulse decay time ($\tau_\mathrm{decay}$)}
The observed burst pulse profiles are well modelled by a Gaussian profile. The Gaussian function gave better $\chi^2$ fits than either the complete analytical profile function derived by AM (see Eq.\,65 in AM) or its non-Gaussian-limiting functional forms. 
Meanwhile, at frequencies above 130\,MHz, a skewed burst profile starts becoming evident. 
However, since the burst pulses are only a few seconds wide, with the rising and decay phases lasting for barely a second or two, it was impossible to get reliable $\chi^2$ fits for separate exponential models in the two phases.
  \begin{figure}
      \centering
      \includegraphics[width=0.4\textwidth,height=6cm]{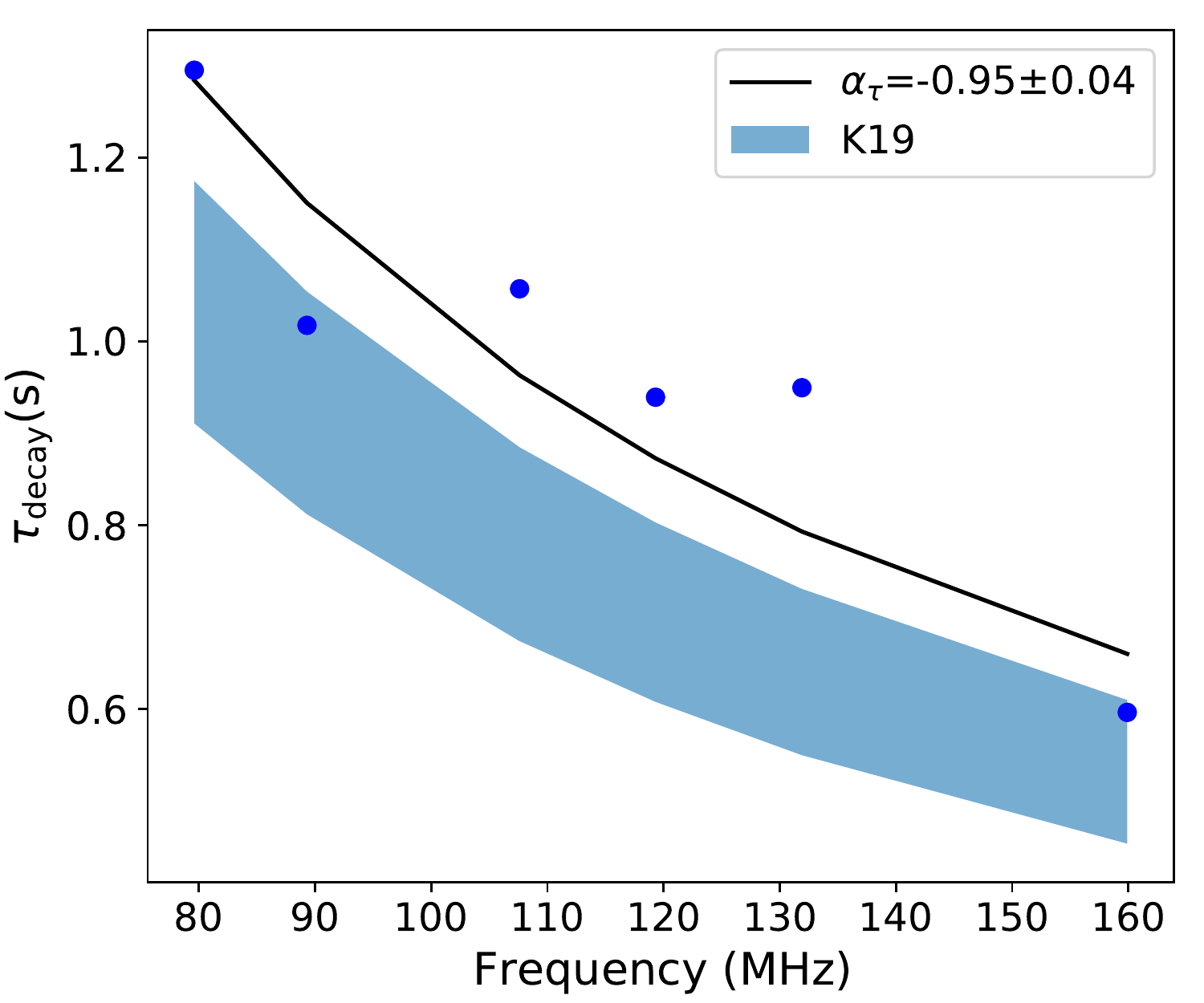}
      \caption{Pulse decay time, $\tau_\mathrm{decay}$, versus frequency. The solid line shows the best-fit power-law function to the data with spectral index, $\alpha_{\tau}$. The shaded band represents the expected range of values from the K19 results.}
      \label{fig8:burstpulse_Vs_nu}
  \end{figure}
Thus, Gaussian profiles were fit to burst pulses across all chosen observation frequencies.
The half width at half maximum of the Gaussian pulse profile gave an estimate of the burst decay time, $\tau_\mathrm{decay}$, across frequencies.
Figure~\ref{fig8:burstpulse_Vs_nu} shows $\tau_\mathrm{decay}(\nu)$ { and the best-fit power-law model with spectral index, $\alpha_{\tau}$, of -0.95$\pm$0.04 as a solid line}. 
Fitting a power law to the archival estimates of $\tau_\mathrm{decay}$ across frequencies, K19 had found a similar spectral index of -0.97$\pm$0.3.
Meanwhile, a comparison of the observed $\tau_\mathrm{decay}$ values with the expected range from the K19 model (shaded band in Fig.~\ref{fig8:burstpulse_Vs_nu}) shows that the absolute values of the observed $\tau_\mathrm{decay}$ are higher than expected.

\subsection{Intrinsic source sizes}
Since the \dnn\ values across the explored heights are well above the respective \dnsat\ values, the region of the corona under study is in the strong turbulence regime.
Therefore, the scatter broadening effect is expected to cause an area rise by an amount equal to the theoretical saturation area ($\pi$\sigsat$^2$). 
The intrinsic source sizes, $\sigma_\mathrm{src}$, can hence be found from the observed source area, $A_\mathrm{obs}$ (Fig.~\ref{fig2:SPREDS}), as follows,
\begin{equation}
    \sigma_\mathrm{src} = \sqrt{\mathrm{A_{obs}/\pi}-\sigma_\mathrm{sat}^2} \label{eqn15:sig_src}
.\end{equation}
Since this equation will be valid only after the source area growth phase, I applied it to the maximum source sizes observed at each frequency. 
  \begin{figure}
      \centering
      \includegraphics[width=0.4\textwidth,height=6cm]{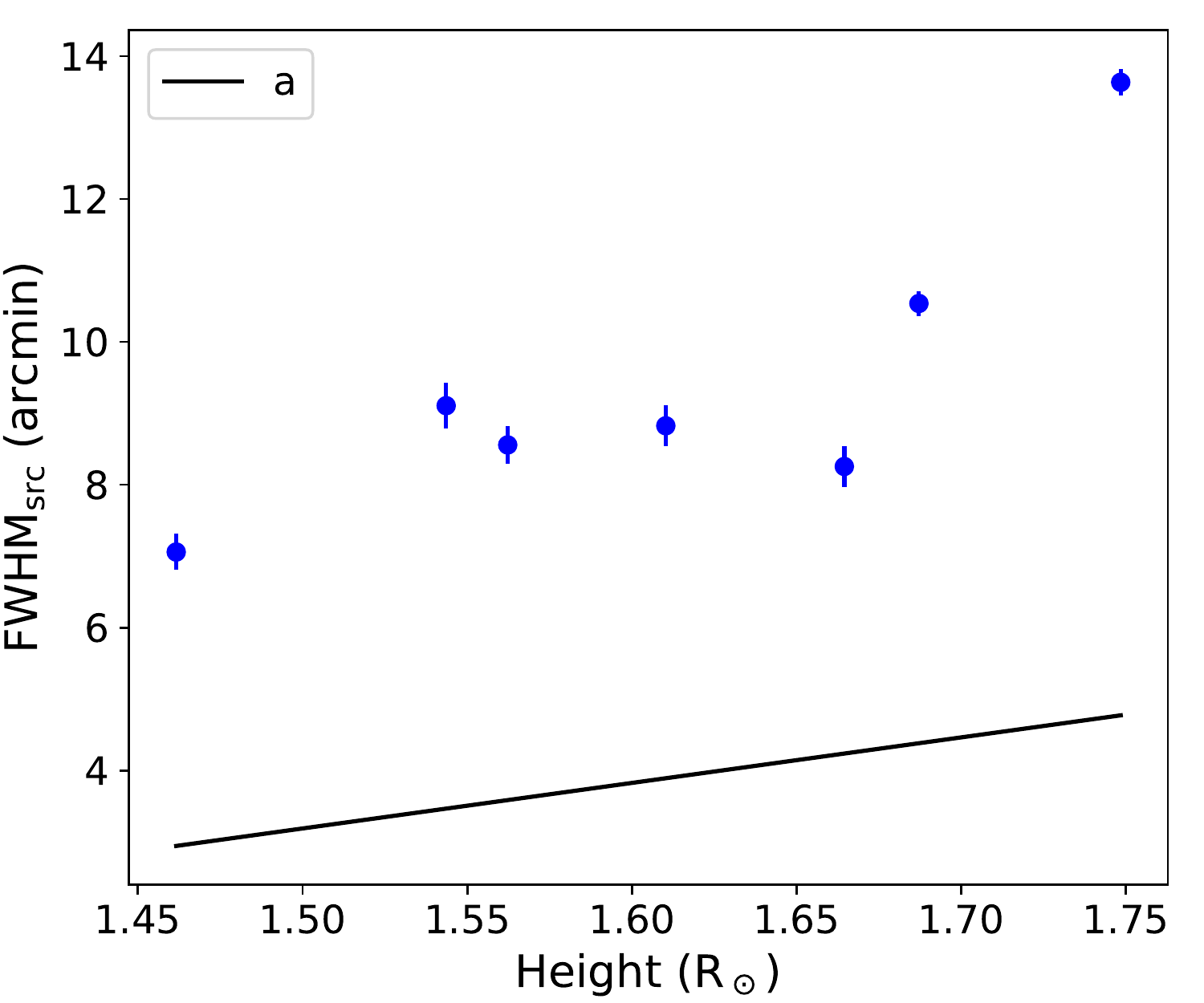}
      \caption{True source sizes (FWHM$_{src}$) versus height after correcting for scatter broadening. The solid line shows the empirical model for the size (a) of coronal flux tubes.}
      \label{fig9:FWHM_true_Vs_ht}
  \end{figure}
Figure~\ref{fig9:FWHM_true_Vs_ht} shows the FWHM values for the intrinsic Gaussian burst source across height (FWHM$_\mathrm{src}$). It is compared with the typical width of magnetic flux tubes (a) across height as per the empirical scaling relation (a$\approx$0.4$\times$[R-1]) proposed by \cite{Asch_observersview2003}. 
Interestingly, the empirical estimates proposed for closed coronal loops are similar to the derived intrinsic source sizes (i.e. within an order of magnitude). The type-III bursts are expected to be associated with open coronal loops \citep{Reid2014}, which are usually wider than the closed loops at a given height.

\section{Conclusion}
\label{sec:conclusions}
I present a snapshot spectroscopic study of the evolution of a type-III burst at sub-megahertz and sub-second resolution in the 80 -- 200\,MHz band. The event was selected from the archival MWA Phase-I database such that it had a well-discernible pulse profile and its source was close to the disk centre { with no nearby bright emission regions}. These criteria ensured that the response of the corona to a burst pulse could be studied in both dynamic spectral and image planes without contamination from co-temporal and co-spatial events. The burst source had a 2D Gaussian structure at all times across frequencies. 
Therefore, 2D Gaussian functions were fitted to the sources and the best-fit parameters were estimated after deconvolving the effect of the synthesised beam. 
The integrated flux density and the area of the beam-deconvolved source was studied across frequencies and time. 

The source area showed a linear rise simultaneous to that in flux density, as expected from a diffusive propagation model of { radio waves subject to strong scattering across the corona.} The area and integrated flux density profiles were modelled using the framework by AM and K19  to derive the strength of density fluctuations (\dnn) across height. The \dnn\ values ranged within 0.4 -- 0.6\% in the observation band. { Combining this with the earlier results in the 30 -- 80\,MHz band, it is found that the \dnn\ values agree within a few factors across 30 -- 140\,MHz ($\approx$ 1.5 -- 2.2\Rsun) despite the differences in the spatial profiles of coronal density and scattering scales assumed by various authors.}
The \dnn\ values roughly scaled with height (R) as R$^{2.2\pm 0.73}$ in the range R $\approx$ 1.54 -- 1.75\,R$_\odot$. The corona under study is found to be in a `strong turbulence regime'. The scattering screen widths derived for the height range are significantly smaller than the outer scale of the density fluctuations { and the different physical property variation scales in the medium}.

The FWHM of the Gaussian source versus frequency was compared against the expected range from the fit derived by K19 using archival observations. It is found that the observed mean and maximum FWHM values roughly cover the expected range. However, the spectral index obtained for the mean and maximum FWHM of the source are -1.0$\pm$0.002 and -0.8$\pm$0.006, which flank the K19 value of -0.98$\pm$0.05. The source sizes showed a spike around 130\,MHz, which correlated with an increased \dnn\ around 1.65 -- 1.7\,R$_\odot$.

The burst decay times ($\tau_\mathrm{decay}$) across frequency also showed a power law with an index of -0.95$\pm$0.04. This is close to the estimate by K19, though the values of $\tau_\mathrm{decay}$ lie outside the range expected from their model based on archival data.

Having estimated the effect of scatter broadening, I derive the intrinsic source sizes across R. The sizes increased with R and agreed with the empirical model for the closed flux tube widths proposed by \cite{Asch_observersview2003} within an order of magnitude.

Apart from the effect of scattering seen in the evolution of the burst source, I report an anti-phased pulsation with a time lag of $\approx$ 2 -- 3\,s in the source area and integrated flux density. This could be attributed to a sausage-mode-like motion at the particle acceleration site that generated the electron beams which in turn produced the type-III bursts.

This work demonstrates the power of snapshot spectroscopic imaging across a wide spectral band for exploring the evolution of coronal turbulence and deriving its characteristics at coronal heights much closer to the Sun than explored by IPS measurements. More of such work in different spectral bands is needed to understand the nature of solar coronal turbulence across a larger height range during different levels of solar activity.

\begin{acknowledgements}
We acknowledge the Wajarri Yamatji people as the traditional owners of the Observatory site. 
Support for the operation of the MWA is provided by the Australian Government's National Collaborative Research Infrastructure Strategy (NCRIS), under a contract to Curtin University administered by Astronomy Australia Limited. We acknowledge the Pawsey Supercomputing Centre, which is supported by the Western Australian and Australian Governments. 
This work is supported by the Research Council of Norway through its Centres of Excellence scheme, project number 262622 (``Rosseland Centre for Solar Physics'').  
AM acknowledges support from the EMISSA project funded by the Research Council of Norway (project number 286853). This research made use of NASA's Astrophysics Data System (ADS). AM is grateful to the developers of Python3 and various packages namely Numpy \citep{numpy}, Astropy \citep{2013A&A...558A..33A}, Scipy \citep{scipy} and Matplotlib \citep{matplotlib}.
\end{acknowledgements}

\bibliographystyle{aa}
\bibliography{allref}
%
%

\end{document}